\documentclass{ws-ijmpa}
\usepackage[super,compress]{cite}
\usepackage{amsmath,amssymb,slashed}
\usepackage{graphicx,xcolor}
\usepackage{hyperref}
\usepackage{placeins}

\begin{document}
	
\title{Properties of size-dependent models having quasiperiodic boundary conditions}
\author{E. Cavalcanti}
\address{Centro Brasileiro de Pesquisas F\'{\i}sicas/MCTI\\
	Rio de Janeiro, RJ, 22290-180, Brazil\\
	erich@cbpf.br}
\author{C.A. Linhares}
\address{Instituto de F\'{\i}sica, Universidade do Estado do Rio
	de Janeiro\\
	Rio de Janeiro, RJ, 20559-900, Brazil\\
	linharescesar@gmail.com}
\author{A. P. C. Malbouisson}
\address{Centro Brasileiro de Pesquisas F\'{\i}sicas/MCTI\\
	Rio de Janeiro, RJ, 22290-180, Brazil\\
	adolfo@cbpf.br}

\maketitle

\begin{history}
	\received{Day Month Year}
	\revised{Day Month Year}
\end{history}

\begin{abstract}
	Boundary conditions effects are explored for size-dependent models in thermal equilibrium. Scalar and fermionic models are used for $D=1+3$ (films), $D=1+2$ (hollow cylinder) and $D=1+1$ (ring). For all models a minimal length is found, below which no thermally-induced phase transition occurs. Using quasiperiodic boundary condition controlled by a contour parameter $\theta$ ($\theta=0$ is a periodic boundary condition and $\theta=1$ is an antiperiodic condition) it results that the minimal length depends directly on the value of $\theta$. It is also argued that this parameter can be associated to an Aharonov-Bohm phase.
\end{abstract}
\ccode{PACS numbers: 11.30.Qc; 11.10.Wx; 11.10.Kk; 03.65 Ta}
	
	\section{Introduction \label{Sec:Intro}}
	The question of the influence of the size of a system is of importance in many situations: e.g., consequences on the transition temperature for systems having some dimensions of finite size, as films, wires, and grains in condensed matter; also in higher dimensional systems with some compactified dimensions.
	
	In previous works,~\cite{Abreu:2003zz,Abreu:2003es,Linhares:2006my,linhares2007critical,Abreu:2009zz,Abreu:2011rj,Abreu:2011zzc,Linhares:2011nh,Linhares:2012vr,linhares2012first,Khanna:2012zz,Khanna:2012js,Abreu:2013nca,Correa:2013mta} when investigating phase transitions in films, periodic or antiperiodic boundary conditions for spatial coordinates have been used in analogy with the imposed condition on the imaginary-time variable. According to the KMS condition,~\cite{TheBook} the boundary conditions on imaginary time are restricted to be periodic for bosons and antiperiodic for fermions.  However, there are no similar restrictions on the spatial boundary conditions. As a generalization, we can study a whole new class of models whose spatial boundary conditions are between the perfect periodic and the perfect antiperiodic boundary conditions, which is a way of generalizing the boundary conditions within the framework of QFT on spaces with toroidal topologies. Although we choose to refer to this as a quasiperiodic boundary condition, it follows along lines similar to those used in \textit{anyonic} systems~\cite{BoschiFilho:1994an,Huang:1994npb,Chen:1995prb,Khare:1996pla,Liguori:2000npb,LeClair:2005JHEP,Shech:2015fph,Lerda} and models with twisted boundary conditions~\cite{Banados:1998ta,Antoniadis:1987wp,Alday:2005ww,Billo:2008sp,Destri:1994bv,Bernard:2010fp,Sachrajda:2004mi,Flynn:2005in,Misumi:2014raa,Petkova:2000ip,Toms:1979ik} and have been found to be useful in many fields, e.g., the Casimir effect~\cite{BoschiFilho:1994an,Flachi:2009tg,Toms:1979ik,Elizalde:2013pfa}, condensed matter systems~\cite{Poilblanc:1991prb,Ryu:2007prb,Bellucci:2010xd}, string theories~\cite{Antoniadis:1987wp,Alday:2005ww,Billo:2008sp}, and also effective and phenomenological models for high-energy physics~\cite{Destri:1994bv,Bernard:2010fp,Sachrajda:2004mi,Flynn:2005in,Misumi:2014raa}.
	
	The study of thermal phase transitions in a quantum field theoretical approach is usually done through either the imaginary-time Matsubara formalism~\cite{Matsubara:1955ws} or the real-time formalism~\cite{Landsman:1986uw}. Throughout this article, we use an extension of the imaginary-time formalism. We consider a $D$-dimensional Euclidean space and introduce periodic/antiperiodic boundary conditions on $d$ of its coordinates, effectively compactifying them, and generating a toroidal topology $\Gamma_D^d = (\mathbb{S}^1)^d \times \mathbb{R}^{D-d}$, with $1\le d\le D$. This defines the so-called quantum field theory on toroidal topologies~\cite{Khanna:2014qqa} which has been applied in the recent literature~\cite{Abreu:2003zz,Abreu:2003es,Linhares:2006my,linhares2007critical,Abreu:2009zz,Abreu:2011rj,Abreu:2011zzc,Linhares:2011nh,Linhares:2012vr,linhares2012first,Khanna:2012zz,Khanna:2012js,Abreu:2013nca,Correa:2013mta}.
	
	In this article, the phase transition for these models is studied by constructing and analyzing the effective potential of the theory through the 2PI formalism. The existence of a  nontrivial minimum of the effective potential corresponds in this case to a phase transition and defines for some models a criticality condition. For instance, for models undergoing a second-order phase transition the criticality is related to a vanishing effective mass.
	
	Before approaching specific problems we present the general formalism for a scalar field in Sec.~\ref{Sec:Formalism} and study its general consequences. Then we apply the formalism both for a scalar and a fermionic model. In Sec.~\ref{Sec:Scalar} we present a scalar model, which is of the Ginzburg-Landau type, and consider some special cases which allow to take into account first-order and second-order phase transitions. The fermionic model is introduced in Sec.~\ref{Sec:Fermion}. The results are presented throughout the article and are synthesized in the conclusions, Sec. \ref{Sec:Discussion}.
	
	We emphasize that we are dealing with phase transitions from a purely theoretical point of view. We are not directly concerned with comparison with experiments. In fact, we are mainly concerned with the mathematical consistency of our approach. Quasiperiodic boundary conditions are similar to \textit{anyonic} statistics largely used over the last years, in connection in particular with the quantum Hall effect.
	Here, differently, we are interested in phase transitions occurring in systems obeying quasiperiodic boundary conditions from a mathematical physics point of view. However, in Sec.~\ref{Sec:Discussion}, we present a discussion in which we interpret the contour parameter as related to an Aharonov-Bohm phase.
	
	\section{The formalism \label{Sec:Formalism}}
	
	Let us take a generic field $\Phi$ for which the boundary condition imposed on the $x_i$ spatial variable is
	\begin{equation}
		\Phi (\ldots, x_i + L, \ldots) = e^{i \pi\theta_i} \Phi(\ldots, x_i, \ldots);
	\end{equation}
	\noindent where $\theta_i=0$ corresponds to a periodic condition and $\theta_i=1$ to an antiperiodic condition. The parameter $\theta_i$ is called the \textit{boundary parameter}. Mathematically, the only change in the general formalism is that the frequencies associated with the spatial compactification become
	\begin{equation}
		p_i \rightarrow \omega_n^i =  \frac{2\pi n_i}{L_i}+ \frac{\theta_i\pi}{L_i}.
		\label{Eq:Identification}
	\end{equation}
	This feature can be absorbed into the formalism by introducing an \textit{imaginary chemical potential} that takes into account the quasiperiodic boundary conditions. We then write
	\begin{equation}
		\omega_n^i = \frac{2\pi n_i}{L_i};\quad\quad
		\mu_i = i \frac{\theta_i\pi}{L_i}.
	\end{equation}
	
	The following integral,
	\begin{equation}
		\mathcal{I}_\nu^D(M^2) = \int \frac{d^D p}{(2\pi)^D} \frac{1}{(p^2+M^2)^\nu},
		\label{Eq:Integral}
	\end{equation}
	\noindent plays an important role in the formalism we develop. It is to be evaluated on a $D$-dimensional Euclidean space, with $M^2$ being the squared field mass. By introducing periodic, antiperiodic or quasiperiodic boundary conditions on $d$ coordinates we effectively map our theory from a Euclidean space ($\mathbb{R}^D$) onto a \textit{toroidal space} ($(\mathbb{S}^1)^d\times\mathbb{R}^{D-d}$). 
	The compactification of imaginary time introduces the temperature $T = \beta^{-1} = L_0$ and compactifications of the spatial coordinates introduce characteristic lengths $L_i$. We apply periodic or antiperiodic conditions to imaginary time if the model is, respectively, bosonic or fermionic, and apply the quasiperiodic boundary conditions to the compactified spatial coordinates. By using a condensed notation in which $i=0$ is associated to the imaginary time, and computing the remaining integral on the $(D-d)$-dimensional subspace using dimensional regularization~\cite{Bollini:1972ui,tHooft:1972tcz}, we get
	\begin{multline*}
		\mathcal{I}_\nu^D(M^2; L_\alpha,\mu_\alpha,\theta_\alpha)= \frac{\sum_{n_0,\ldots,n_{d-1}=-\infty}^{\infty}}{\prod_{\alpha=0}^{d-1} L_\alpha}\int \frac{d^{D-d} q}{(2\pi)^{D-d}} \frac{1}{[q^2 + M^2 + \sum_{\alpha=0}^{d-1} (\frac{2\pi n_\alpha}{L_\alpha} - i \mu_\alpha)^2]^\nu}\\
		= \frac{\sum_{n_0,\ldots,n_{d-1}=-\infty}^{\infty}}{\prod_{\alpha=0}^{d-1} L_\alpha} \frac{\Gamma[\nu-\frac{D}{2}+\frac{d}{2}]}{(4\pi)^{\frac{D}{2}-\frac{d}{2}}\Gamma[\nu]} \frac{1}{[M^2+\sum_{\alpha=0}^{d-1} (\frac{2\pi n_\alpha}{L_\alpha} - i \mu_\alpha)^2 ]^{\nu-\frac{D}{2}+\frac{d}{2}}},
	\end{multline*}
	\noindent where $\Gamma(\nu)$ is the Euler gamma function. In the above formula the summations over $n_0$ and $\{n_i\}$ correspond to compactification of, respectively, the imaginary time and the spatial coordinates.
	
	We define the quantity $\mu_0 = \mu + i\theta_0 \pi/\beta$, which depends on the chemical potential $\mu$ and also on whether the model describes bosons ($\theta_0=0$) or fermions ($\theta_0=1$). The remaining infinite sum can be identified as an Epstein-Hurwitz zeta function, which can be regularized by the use of a Jacobi identity for theta functions, leading to sums of modified Bessel functions of the second kind $K_\nu(X)$ (see Ref.~\cite{Elizalde:1996zk}),
	\begin{multline} 
		\mathcal{I}_\nu^D(M^2; L_\alpha,\mu_\alpha,\theta_\alpha) = 
		\frac{(M^2)^{-\nu+\frac{D}{2}} \Gamma\left[\nu-\frac{D}{2}\right]}{(4\pi)^{\frac{D}{2}}\Gamma[\nu]}
		+ \frac{1}{(2\pi)^{\frac{D}{2}} 2^{\nu-2}\Gamma[\nu]} \times\\\times
		\Bigg\{
		\sum_{\alpha=0}^{d-1}\sum_{n_\alpha=1}^\infty \left(\frac{n_\alpha L_\alpha}{M}\right)^{\nu-\frac{D}{2}} \cosh(n_\alpha L_\alpha \mu^\alpha) K_{\nu-\frac{D}{2}}(n_\alpha L_\alpha M)
		+ \cdots \\+ 2^{d-1}\hspace{-17pt}\sum_{n_0,\ldots,n_{d-1}=1}^\infty\hspace{-5pt} \left(\hspace{-5pt}\frac{\sqrt{\sum_{\alpha=0}^{d-1}n_\alpha^2 L_\alpha^2}}{M}\right)^{\nu-\frac{D}{2}} \hspace{-5pt}\prod_{\alpha=0}^{d-1}\cosh(n_\alpha L \mu^\alpha) K_{\nu-\frac{D}{2}}\left(\hspace{-5pt}M\sqrt{\sum_{\alpha=0}^{d-1}n_\alpha^2 L_\alpha^2}\right) 
	\hspace{-5pt}	\Bigg\}.\label{funcInD}
	\end{multline}
	
	In the following we restrict ourselves to the $d=2$ case, so that the compactifications introduce the temperature $L_0=\beta^{-1}$; a characteristic length $L_1=L$; the parameter $\mu_0=\mu+i \theta_0\pi/\beta$, which carries information about the chemical potential $\mu$ and the imaginary-time boundary condition; and the parameter $\mu_1=i\theta\pi/L$ which carries information about the spatial quasiperiodic boundary condition.
	
	The function $\mathcal{I}_\nu^D$ in Eq.~\eqref{funcInD} can be rewritten as
		\begin{equation}
			\mathcal{I}_\nu^D(M^2; \beta,\mu,\theta_0;L,\theta)
			= \frac{(M^2)^{-\nu+\frac{D}{2}} \Gamma\left[\nu-\frac{D}{2}\right]}{(4\pi)^{\frac{D}{2}}\Gamma[\nu]} + \frac{W_{\frac{D}{2}-\nu} \left[M^2; \beta,\mu,\theta_0;L,\theta\right]}{(2\pi)^{\frac{D}{2}} 2^{\nu-2}\Gamma[\nu]},
			\label{Eq:GenericProp}
		\end{equation}
		\noindent where the function $W_\rho$, introduced to simplify notations, is defined by
		\begin{multline}
			W_\rho\left[M^2; \beta,\mu,\theta_0;L,\theta\right] = \sum_{n=1}^\infty\Bigg\{ \left(\frac{M}{n \beta}\right)^\rho (-1)^{n\theta_0}\cosh(n \beta \mu) K_\rho(n \beta M)
			+ \left(\frac{M}{n L}\right)^\rho \cos(n \pi \theta) K_\rho(n L M)\Bigg\}\\
			+ 2 \sum_{n_0,n_1=1}^{\infty} \frac{M^\rho (-1)^{n_0\theta_0} \cosh(n_0 \beta \mu) \cos(n_1 \pi \theta)}{\left(n_0^2 \beta^2+n_1^2L^2\right)^{\rho/2}} K_\rho\left(M \sqrt{n_0^2 \beta^2+n_1^2L^2}\right),
			\label{Eq:FunctionW}
		\end{multline}
	\noindent which is positive and monotonically decreasing with $L$ and $\beta$. Its derivatives are computed by means of the recurrence formula
	\begin{equation}
		\frac{d^k}{d X^k} W_{\nu}[X,\beta,L,\mu,\theta] = \left(-\frac{1}{2}\right)^k W_{\nu-k}[X,\beta,L,\mu,\theta].
	\end{equation}
	
	With respect to the function $W_\rho$, which contains all size and temperature dependencies, it turns out that an expression for $M^2=0$ is useful in many occasions. For $\rho>0$ we obtain, by taking the modified Bessel function of the second kind in the limit $M\rightarrow0$ and computing the sum over the frequencies,
	\begin{multline*}
		W_\rho\left[0; \beta,\mu,\theta_0;L,\theta\right] = 
		\Gamma[\rho] 2^{\rho-2} \Bigg\{
		\frac{\Re\left[\text{Li}_{2\rho} \left((-1)^{\theta_0}e^{\beta \mu}\right)\right]}{\beta^{2\rho}}
		\\+\frac{\Re\!\left[\text{Li}_{2\rho} \left(e^{i \pi \theta}\right)\right]}{L^{2\rho}}
		+ 4\!\!\!\!\! \sum_{n_0,n_1=1}^{\infty} \!\!\!\frac{(-1)^{n_0\theta_0} \cosh(n_0 \beta \mu) \cos(n_1 \pi \theta)}{\left(n_0^2 \beta^2+n_1^2L^2\right)^{\rho}}\Bigg\},
		\label{Eq:Wpolilog}
	\end{multline*}
	\noindent where $\text{Li}_s$ is the polylogarithm function of order $s$. For $\rho=0$ we have, instead, in the $T=0$ case,
	\begin{equation}
		W_0\left[M\rightarrow 0; L,\theta\right] = \frac{\gamma}{2} + \frac{1}{2}\ln \frac{M L}{2} + \frac{\text{Li}'_0(-e^{-i \pi \theta}) + \text{Li}'_0(-e^{i \pi \theta})}{2}.
		\label{Eq:Wpolilog0}
	\end{equation}
	
	As $\Re\left[\text{Li}_{2\rho} \left(e^{i \pi \theta}\right) \right] = \left[\text{Li}_{2\rho} \left(e^{+ i \pi \theta}\right) + \text{Li}_{2\rho} \left(e^{- i \pi \theta}\right)\right]/2$, it is always possible to define a \textit{critical parameter} $\theta^\star$ for which the polylogarithm function vanishes. Its value depends only on $\alpha$ ($\rho=\frac{D}{2}-\nu$, as in Eq.~\eqref{Eq:GenericProp}). Some of these values are exhibited at Table~\ref{tab:theta}.
	\begin{table}[ph]
	\tbl{Root $\theta^\star$ of the polylogarithm function depending on the parameter $\rho = D/2-\nu$.}
	{\begin{tabular}{|c|c|c|c|c|c|c|c|c|}
		\hline
		$\rho$ & 0 & 1/2 & 1 & 3/2 & 2 & 5/2 & \ldots &$\infty$\\
		\hline
		$\theta^\star$ & 0& 1/3 & 0.422650 & 0.461659 & 0.480670 & 0.490238 &\ldots& 1/2\\
		\hline
	\end{tabular}
	\label{tab:theta}}
	\end{table}
	
	Notice that the maximal possible value for $\theta^\star$ is $1/2$, representing the intermediate point between the periodic and antiperiodic boundary conditions.
	
	\section{Ginzburg-Landau model \label{Sec:Scalar}}
	As a first example, we take a Ginzburg-Landau (GL) model with a 6-th order polynomial potential in $D$-dimensions,
	\begin{align}
		S_E(\phi) = \int d^Dx\; \left[\frac{1}{2}(\partial\phi)^2 + V_0 (\phi)\right];	\label{Eq:ActionGL}\\
		V_0(\phi) = m_0^2 \frac{\phi^2}{2} + \lambda_0 \frac{\phi^4}{4!} + g_0 \frac{\phi^6}{6!}.
	\end{align}
	
\begin{figure}[h]
		\centering
		\includegraphics[width=0.5\linewidth]{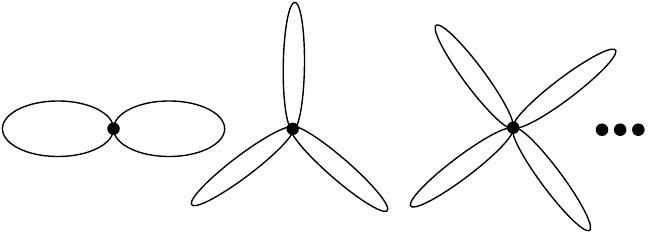}
		\caption{Contributions of the \textit{petal} diagrams.}
		\label{fig:petals}
	\end{figure}

	By employing the 2PI formalism~\cite{Cornwall:1974vz} in the Hartree-Fock approximation, we are restricted to diagrams with only one vertex; then the effective action is written as
	\begin{equation}
		\Gamma_{\text{eff}}(\phi,G) = S_E(\phi) 
		+ \frac{1}{2} [V_0''(\phi)-m^2]\text{Tr}\;G
		+ \frac{1}{2} \text{Tr} \ln G^{-1}
		+ \sum_{n=2}^\infty \frac{V_0^{(2n)}(\phi)}{(2n)!!} (\text{Tr}\;G)^n,
	\end{equation}
	\noindent where $G$ is the full propagator, which depends on the thermal mass, and the trace $\text{Tr}$ is taken over the $D$-dimensional spatial coordinates and momenta. The last term is the sum over all \textit{petal} diagrams in Fig.~\ref{fig:petals}.
	
	We assume that $\phi$ is a constant field and then define the effective potential as the effective action divided by the $D$-dimensional volume. All size and temperature dependencies are contained in $\text{tr}\; G$ (see Sec.~\ref{Sec:Formalism} for a detailed explanation), and the remaining trace $\text{tr}$ is only over the $D$-momenta. The $\phi$-dependent effective potential is simply
	\begin{equation}
		V_{\text{eff}} (\phi)\hspace{-2pt} =\hspace{-2pt} \left[m_0^2 \hspace{-2pt}+\hspace{-2pt} \frac{\lambda_0}{2} \mathcal{I}_1^{D}(m^2; \beta,\mu,0;L,\theta)\right]\hspace{-5pt}\frac{\phi^2}{2}
		\\+ \left[\lambda_0\hspace{-2pt} +\hspace{-2pt} \frac{g_0}{2} \mathcal{I}_1^{D}(m^2; \beta,\mu,0;L,\theta)\right] \hspace{-5pt}\frac{\phi^4}{4!} + g_0 \frac{\phi^6}{6!},
	\end{equation}
	\noindent where $\mathcal{I}_\nu^{D}$ is defined in Eq.~\eqref{Eq:GenericProp}.
	
	The phase transition analysis consists in determining the value $\varphi$ that minimizes $V_{\text{eff}}$. We must use in parallel the relation $\partial V_{\text{eff}}/\partial \phi \Big|_{\phi=\varphi\neq0}=0$, defining extrema, and $\partial^2 V_{\text{eff}}/\partial \phi^2 \Big|_{\phi=\varphi} = m^2$, which is the recurrence equation defining the thermal dependent mass. The symmetric phase $\varphi=0$ is an acceptable extremum and the mass in this phase evolves as
	\begin{equation}
		m_\text{sym}^2 = m_0^2 + \frac{\lambda_0}{2} \mathcal{I}_1^{D}(m_{\text{sym}}^2; \beta,\mu,0;L,\theta). \label{Eq:SymMass}
	\end{equation}
	\noindent Similarly, in the broken phase we can always find a recurrence relation for the properly defined mass $m^2_{\text{brk}}$ by using a non-trivial minimum $\varphi\neq0$.
	
	\subsection{Phase Transitions}
	\subsubsection{Second-Order Phase Transition}
	
	In this section we consider a theory as described in Eq.~\eqref{Eq:ActionGL} with $g_0=0$. Thus, in the absence of the $\phi^6$ coupling, the system undergoes a second-order phase transition at $m^2=0$~\cite{LeBellac}. Considering the mass evolution from the symmetric phase given in Eq. \eqref{Eq:SymMass} the critical condition is
	\begin{equation}
		m_0^2 + \frac{\lambda_0}{2} \mathcal{I}_1^{D}(0; \beta_c,\mu,0;L,\theta) = 0. 
	\end{equation}
	
	\begin{figure}[h]
		\centering
		\includegraphics[width=0.5\linewidth]{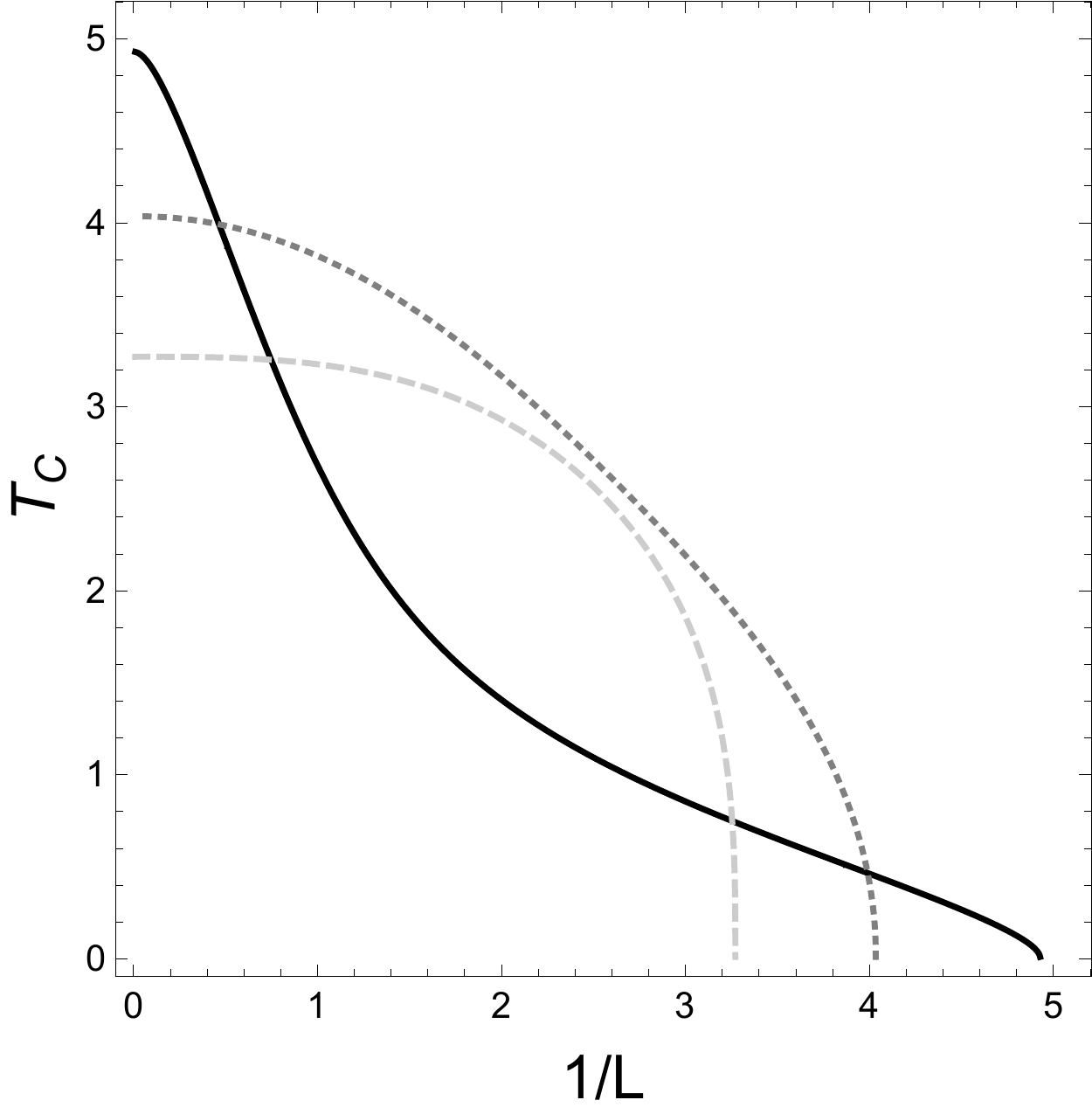}
		\caption{Critical temperature as a function of the inverse length for $D=4$ (full black line), $D=5$ (dashed light gray line), $D=6$ (dotted gray line) for the Ginzburg-Landau model with quartic interaction. These phase diagrams are for the second-order phase transition and exhibits the minimal length (maximal inverse length) below which no phase transition occurs. In each case, the broken phase lies below the respective curve.}
		\label{fig:MinLength}
	\end{figure}
	
	Using this condition we construct a phase diagram giving the critical temperature as a function of the size of the system, in Fig.~\ref{fig:MinLength}, for $D=2,3,4$ and periodic boundary condition $\theta=0$, which exhibits a minimal length for $T_c=0$, below which no thermally-induced phase transition occurs. For systems subject to external influence (for instance an applied magnetic field, pressure,\ldots) phase transitions can occur even for $T=0$, known in the literature as quantum phase transitions~\cite{Muciobook}; however, these situations are beyond the scope of the present work. The behavior in Fig.~\ref{fig:MinLength} is mathematically expected as all size and temperature dependencies are contained in $W_\rho\left[M^2;\beta,\mu,0;L,0\right]$, which is monotonically decreasing in $L$ and $\beta$; to sustain a fixed value for $W$ when $T\rightarrow0$ ($\beta \rightarrow \infty$) the parameter $L$ must decrease. Therefore, for $T_c=0$ the system has its minimal possible length $L_{\text{min}}$ and the critical condition becomes
	\begin{equation}
		0 = m_0^2 + \frac{\lambda_0}{2}
		\Bigg(
		\frac{m^{D-2}\Gamma\left[1-\frac{D}{2}\right]}{(4\pi)^{\frac{D}{2}}}
		+\frac{2}{(2\pi)^\frac{D}{2}}  \frac{\Gamma[\frac{D}{2}-1] 2^{\frac{D}{2}-2}}{L_{\text{min}}^{D-2}} \Re\left[ \text{Li}_{D-2} \left(e^{i \pi \theta}\right)\right]
		\Bigg).
	\end{equation}
	
\begin{figure}[h]
		\centering
		\includegraphics[width=0.5\linewidth]{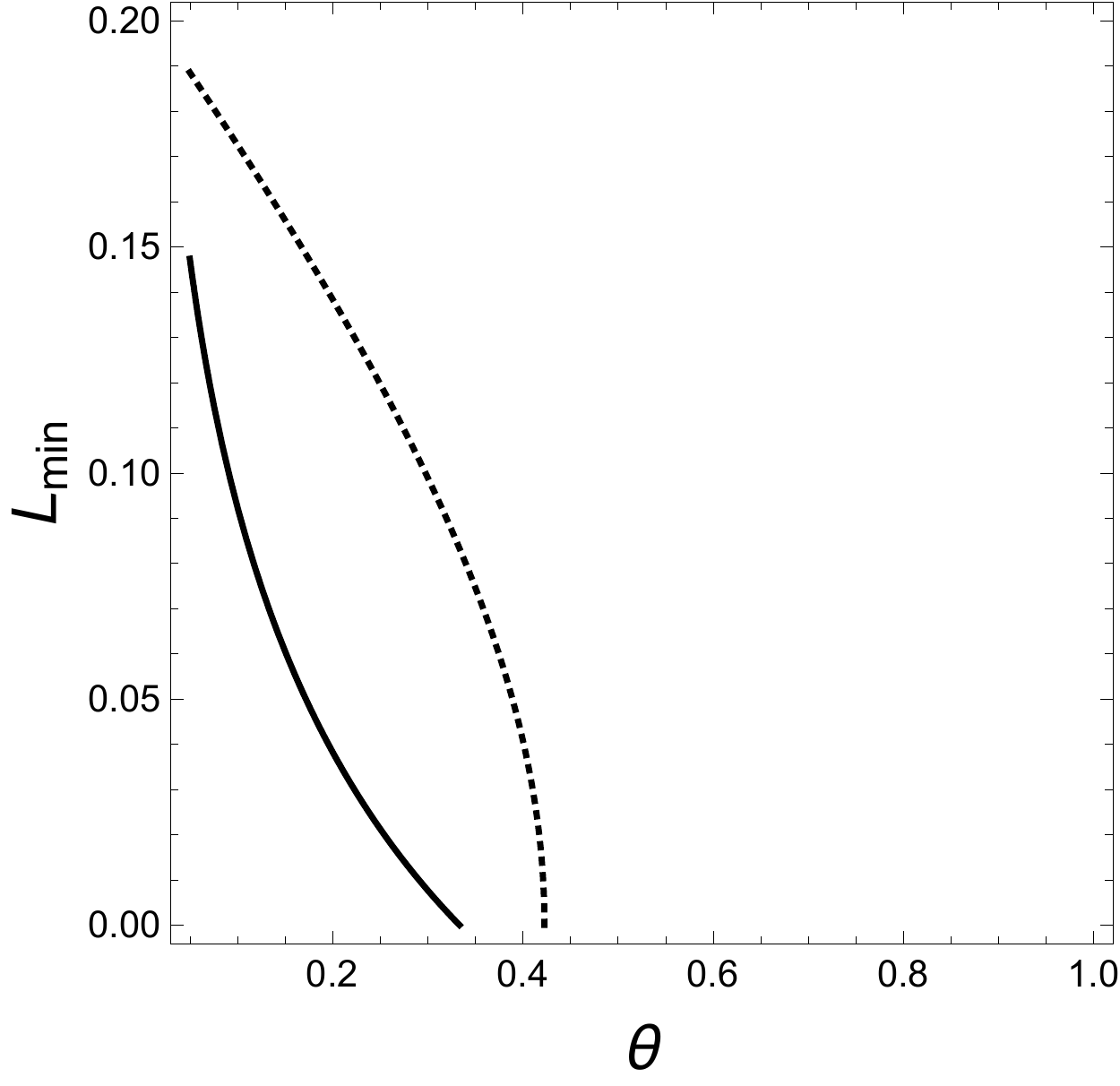}
		\caption{Dependence of the ratio $L_{\text{min}}$ on the contour parameter for $D=1+2$ (full line) and $D=1+3$ (dotted line) in the quartic Ginzburg-Landau model. The region below each curve corresponds to the symmetric phase. The length $L$ has mass dimensions, which is equivalent to take $m_0^2 = -1$, and we are assuming $\lambda=1$.}
		\label{fig:Boson_Lmin}
	\end{figure}

	For some critical value of the contour parameter $\theta=\theta^\star$ the minimal length becomes zero, meaning that the size restriction was removed. The evolution of the minimal length with respect to the contour parameter is presented in Fig.~\ref{fig:Boson_Lmin} for a hollow cylinder ($D=1+2$) and a film ($D=1+3$). Under the critical curve $L_{\text{min}}(\theta)$ a thermally-induced phase transition cannot exist. This justifies the name \textit{minimal length}; under this critical size the system no longer exhibits a phase transition. Above the critical curve $L_{\text{min}}(\theta)$ the phase is broken and there may be a thermally-induced phase transition at some critical temperature.
	
	We emphasize the important contribution of the contour parameter $\theta$ that controls the periodicity; Its value determines whether the system exhibits a minimal length. The critical parameter varies with dimensionality: for $D=2,3,4$ we have respectively $\theta^\star = 0, \frac{1}{3}, 0.42265$. Therefore, the behavior is present for the film model, controlling the minimal film thickness, and for the cylindric model (a tube) controlling its radius. For the ring model the contour parameter has no influence; mathematically this happens because of the property $\text{Li}_{0} (e^{x}) + \text{Li}_{0} (e^{-x}) = -1$; so there is no $\theta$-dependence. This suggests that the contour condition does not modify the minimal radius of a ring ($D=1+1$).
	
\begin{figure}[h]
		\centering
		\includegraphics[width=0.5\linewidth]{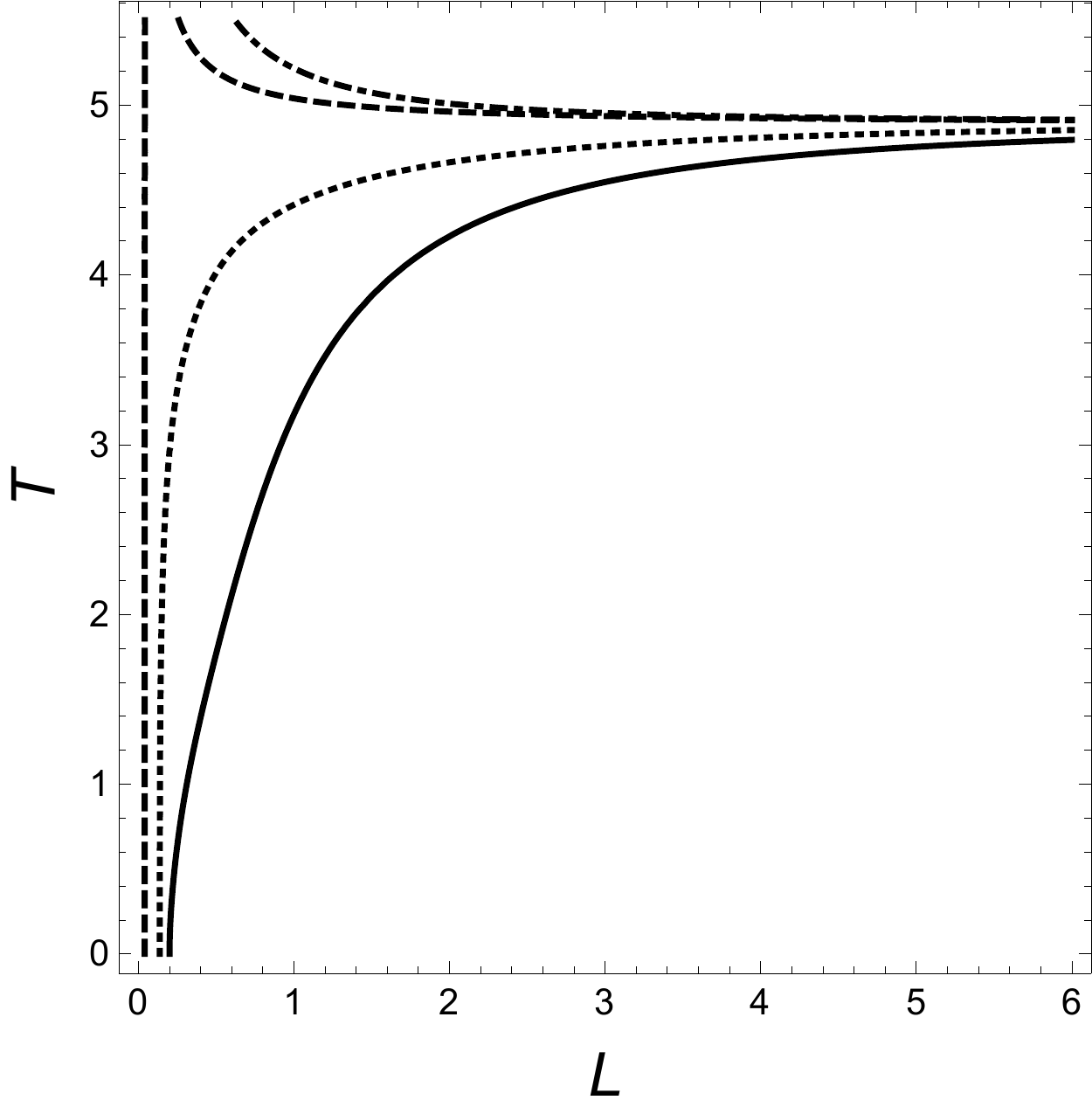}
		\caption{Phase diagram for $D=4$, critical temperature $T_c$ of a bosonic second-order phase transition as a function of the length $L$ for values of the contour parameter $\theta=0,0.2,0.4,0.6$, respectively, the full, dotted, dashed and dotdashed curves. The broken phase is the region below each curve.}
		\label{fig:2ndBound}
	\end{figure}

	For clarity we show in Fig.~\ref{fig:2ndBound} the meaning of a vanishing minimal length: in this case we have $D=4$ and a critical contour parameter $\theta^\star=0.42265$. For $\theta=0,0.2,0.4<\theta^\star$ there is still a minimal length; however, when $\theta=0.6>\theta^\star$ (dot-dashed curve) we no longer have a minimal length and the behavior of the critical temperature as a function of the length is completely changed.
	
	\subsubsection{First-Order Phase Transition}
	
	In this section we consider $g_0\neq 0$, and $\lambda_0<0$. In this case, there is a first-order transition in the GL model of Eq.~\eqref{Eq:ActionGL}. Its critical region is determined by the coexistence $V_{\text{eff}} (\varphi \neq 0) = V_{\text{eff}} (\varphi = 0)$, with $\varphi$ defined as an extremum. From the perspective of the symmetric phase, the critical condition is obtained, after some algebraic manipulations, as
	\begin{equation}
		\mathcal{I}_1^{D}\left(5 m_0^2 - \frac{5 \lambda_0^2}{2 g};\beta_c,\mu,0;L,\theta\right) = - \frac{2\lambda_0}{g} \pm \frac{4\lambda_0}{g}\sqrt{\frac{2 g m_0^2}{\lambda_0^2} - 1}.
		\label{Condition1st}
	\end{equation}
	
\begin{figure}[h]
		\centering
		\includegraphics[width=0.5\linewidth]{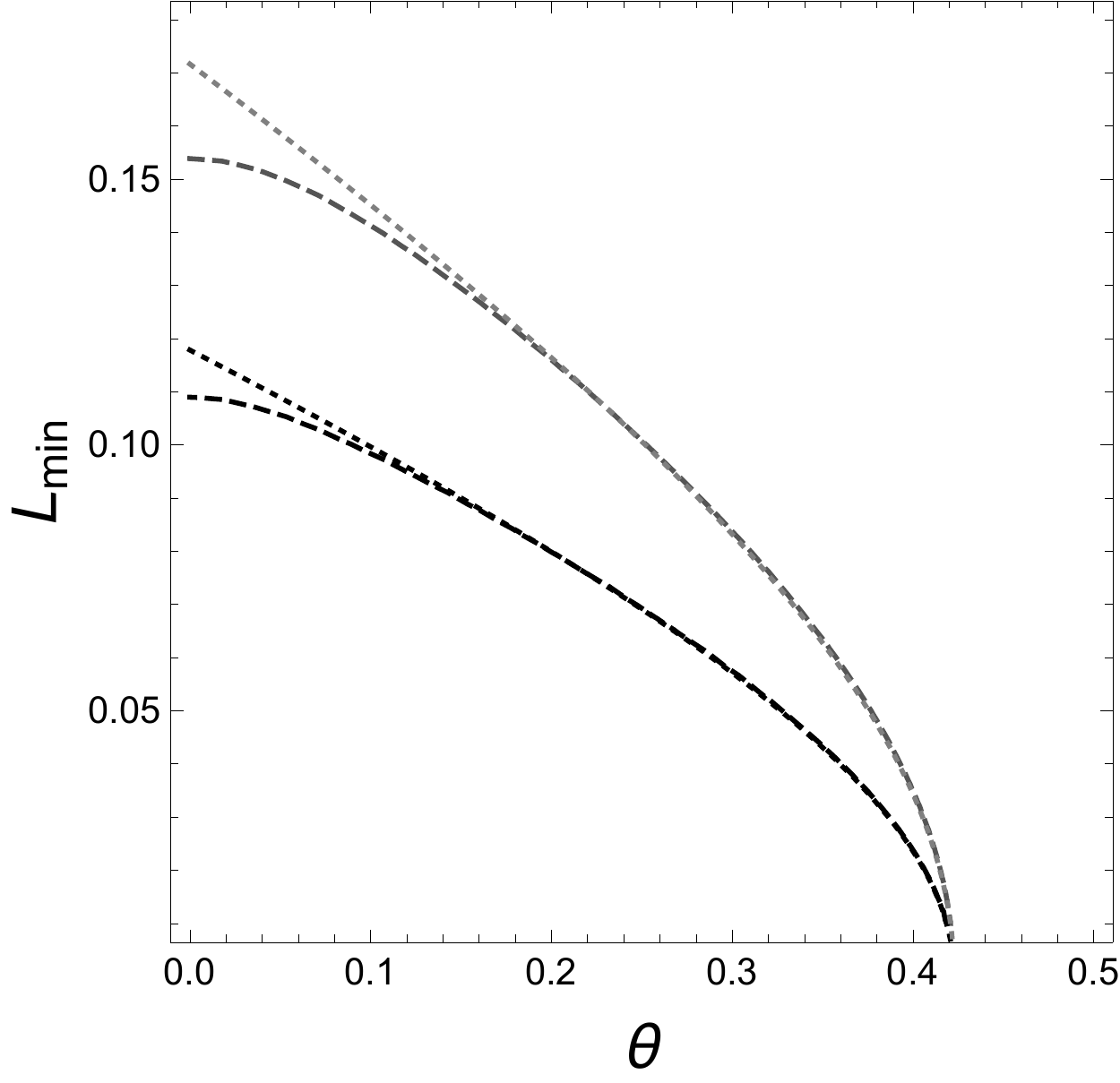}
		\caption{Minimal thickness $L_{\text{min}}$, in $D=1+3$, as a function of the contour parameter $\theta$ in the extended Ginzburg-Landau model. The dashed curve uses the full equation with a truncated series, the dotted curve uses the approximation for low $L_{\text{min}}$. For all curves $m_0^2 = 1$ and $g=1$. The black curves have $\lambda_0 = -1$, implying $g m_0^2/\lambda_0^2 = 1 > 5/8$ which is symmetric at the tree level. The gray curves have $\lambda_0 = - \sqrt{2}$ implying $g m_0^2/\lambda_0^2 = 1/2 < 5/8$ (broken phase at the tree level).}
		\label{fig:1stOrderContPar}
	\end{figure}

	As before, the minimal length $L_{\text{min}}$ is defined as the size of the system at which the critical temperature vanishes ($T_c = 0$) which, as already mentioned, means that there is no thermally induced phase transition for lengths below $L_\text{min}$. Then, by taking this limit we obtain $\lim\limits_{\beta\rightarrow\infty}\mathcal{I}_1^{D}(m^2;\beta,\mu,0;L,\theta) = \mathcal{I}_1^{D}(m^2;L,\theta)$. In this case the condition expressed in Eq.~\eqref{Condition1st} becomes
	\begin{multline}
		\frac{2W_{\frac{D}{2}-1} \left[m^2;L_{\text{min}},\theta\right]}{(2\pi)^{\frac{D}{2}}} = \frac{m^{\frac{D}{2}-1}}{2^{\frac{D}{2}-1}\pi^{\frac{D}{2}} L_{\text{min}}^{\frac{D}{2}-1}} \sum_{n=1}^{\infty} \frac{\cos(n\pi \theta)}{n^{\frac{D}{2}-1}} K_{\frac{D}{2}-1}(n L_{\text{min}} m)\\= - \frac{2\lambda_0}{g} \pm \frac{4\lambda_0}{g}\sqrt{\frac{2 g m_0^2}{\lambda_0^2} - 1}. \label{Eq:Boson1st:Lmin}
	\end{multline}
	As an example, we consider $D=1+3$ (a film) and investigate the critical contour parameter $\theta^\star$ at which there is no minimal length, see Fig.~\ref{fig:1stOrderContPar}. This can be done by taking in Eq.~\eqref{Eq:Boson1st:Lmin} $L_{\text{min}} \approx 0$ and using an asymptotic formula for $K_\nu(z)$ for $z\sim0$, so that,
	\begin{equation}
		L_{\text{min}} = \left(- \frac{2\lambda_0}{g} \pm \frac{4\lambda_0}{g}\sqrt{\frac{2 g m_0^2}{\lambda_0^2} - 1}\right)^{-\frac{1}{2}}\sqrt{\frac{\Re\left[\text{Li}_2(e^{i \pi \theta})\right]}{2\pi^2}}. 
	\end{equation}
	\noindent In this case we obtain the value $\theta^\star = 0.42265$. We see that, although we are dealing with a first-order phase transition, this is the same result of the previous section where we dealt with a second-order phase transition. This means that the critical contour parameter seems to be a natural characteristic of the compactified scalar model, regardless the order of the phase transition.
	
	In Fig.~\ref{fig:1stOrderContPar} we compare the approximation for a low value of $L_\text{min}$ (dotted line) and the full equation (dashed line), note that they only disagree for very low values of $\theta$. The presence of the critical contour parameter at which the minimal length goes to zero is made evident. Let us consider two different initial conditions at the tree level, one ensuring that the phase is symmetric (black lines) and the other one ensuring that the phase is broken (gray lines); both exhibit the same behavior when taking $T_c=0$ and varying the minimal length with respect to the contour parameter.
	
	\section{The Bosonized Gross-Neveu model \label{Sec:Fermion}}
	
	In this section we extend the massless Gross-Neveu model originally established for $D=1+1$~\cite{Gross:1974jv}, to generic $D$ dimensions where the model is not renormalizable but can be viewed under some circumstances as an effective model for QCD~\cite{Khanna:2012zz,Khanna:2012js}. In this case, perturbative renormalizability is not an absolute criterion for the existence of the model~\cite{Parisi:1975im,Gawedzki:1985ed,Gawedzki:1985jn}. We point out that for $D=1+2$, although not perturbatively renormalizable the model has been shown to exist and was constructed~\cite{deCalan:1991km}. We take into account temperature, chemical potential, finite-size effects and the contour condition. Both the fermion ring (1+1) and the fermion tube (1+2) are constructed by identifying a space point (compactifying the space) which modifies its topology; this compactification is controlled by the contour parameter. We find that the system exhibits a dynamical generation of mass~\cite{Miransky} that here characterizes a second-order phase transition. A minimal length below which no thermally induced phase transition occurs is found in both cases, which means that the fermion ring does not become a point and that the fermion tube does not become a line and both have dependencies on the contour parameter.
	
	We consider a colourless and flavourless fermionic system with an interaction of the Gross-Neveu type, 
	\begin{equation}
		S(\bar{\psi},\psi) = \int d^Dx\; \left[\bar\psi \slashed \partial \psi + g_0^2 (\bar\psi \psi)^2\right].
	\end{equation}
	\noindent In our convention we use Euclidean $\gamma$ matrices~\cite{zinnjustin2002}.
	We consider the bosonization given by the scalar field $\sigma = \bar{\psi} \psi$. To find the new Lagrangian density we then employ the substitution $(\bar \psi \psi)^2 = 2 \bar \psi \psi \sigma - \sigma^2$, which ensures that the relation $\delta S / \delta \sigma = 0$ leads to the identity $\sigma=\bar \psi \psi$. We then obtain that the action is 
	\begin{equation}
		S(\bar{\psi},\psi,\sigma) = \int d^Dx\; \left[\bar\psi 
		\left(\slashed \partial + g_0^2 \sigma \right)
		\psi 
		- \frac{g^2_0}{2} \sigma^2
		\right],
	\end{equation}
	\noindent and the generating function is 
	\begin{equation*}
	Z = \int \mathcal{D}[\bar \psi, \psi, \sigma] e^{-S(\bar{\psi},\psi,\sigma)}. 
	\end{equation*}
	\noindent Integrating over the fermionic field, we construct the effective potential,
	\begin{equation*}
		V_{\text{eff}}(\sigma) = \frac{g^2_0}{2} \sigma^2 - \text{Tr} \ln \left[\slashed \partial + g^2_0 \sigma  \right],
	\end{equation*}
	\noindent where the trace is to be evaluated over the Dirac indices and the momentum space. Using that $\text{Tr} \ln = \ln \text{Det}$ and taking the determinant over the Dirac indices, we get
	\begin{equation*}
		V_{\text{eff}} = \frac{g^2_0}{2} \sigma^2 - \int \frac{d^Dp}{(2\pi)^D} \ln [p^2 + g^4_0 \sigma ^2].
	\end{equation*}
	\noindent The logarithm can be expressed as a derivative
	\begin{equation*}
		\ln x = - \frac{\partial}{\partial \nu} x^{-\nu} \Bigg|_{\nu=0},
	\end{equation*}
	\noindent which allows us to employ Eq.~\eqref{Eq:GenericProp} and then obtain
	\begin{equation*}
		V_{\text{eff}}=
		\frac{g^2_0}{2} \sigma^2
		+ \frac{\Gamma\left[-\frac{D}{2}\right]}{(4\pi)^\frac{D}{2}} (g^2_0 \sigma )^D
		+ \frac{4}{(2\pi)^\frac{D}{2}}W_{\frac{D}{2}} \left[(g^2_0 \sigma)^2;\beta,\mu,1;L,\theta\right],
	\end{equation*}
	\noindent where we have used that $$\frac{\partial}{\partial \nu} \frac{f(\nu)}{\Gamma(\nu)} \Big|_{\nu=0} = f(0)$$ for a function $f(\nu)$ with no poles at $\nu=0$. The function $W_\nu$ was defined in Eq. \eqref{Eq:FunctionW}. The term $\theta_0=1$ corresponds to the antiperiodic boundary condition on the imaginary time, which is used since we are dealing with a fermionic model.
	
	The dynamically generated mass is $m = g_0^2 \sigma$, so we can rewrite the effective potential as
	\begin{equation*}
		V_{\text{eff}}=
		\frac{m^2}{2g^2_0}
		+ \frac{\Gamma\left[-\frac{D}{2}\right]}{(4\pi)^\frac{D}{2}} |m|^D
		+ \frac{4}{(2\pi)^\frac{D}{2}}W_{\frac{D}{2}} \left[m^2;\beta,\mu,1;L,\theta\right].
	\end{equation*}
	\noindent By applying the renormalization condition
	\begin{equation}
		\frac{\partial^2 V_{\text{eff}}}{\partial m^2} (m=m_R;\beta\rightarrow\infty) = \frac{1}{g_R},
	\end{equation}
	\noindent we exchange the effective potential dependence from $g_0$ and $m$ to $g_R$ and $m_R$, leading to
	\begin{equation*}
		V_{\text{eff}}=
		\frac{m^2}{2g^2_R}
		+\frac{\Gamma\left[-\frac{D}{2}\right]}{(4\pi)^\frac{D}{2}}\left(|m|^D-
		\frac{m^2D (D-1)}{2}|m_R|^{D-2}\right)
		+ \frac{4}{(2\pi)^\frac{D}{2}}W_{\frac{D}{2}} \left[m^2;\beta,\mu,1;L,\theta\right].
	\end{equation*}
	\noindent Alternatively we express the effective potential in terms of the dynamically generated mass defined by the condition $\partial V_{\text{eff}}/\partial m \Big|_{m=\widetilde m} = 0$, taking the point at zero temperature $\widetilde{m}(T=0,\mu=0,\frac{1}{L}=0) = m_0$. Then the effective potential is written as
	\begin{multline}
		V_{\text{eff}}=
		m^2 \frac{\Gamma\left[-\frac{D}{2}\right]}{(4\pi)^\frac{D}{2}} 
		\left(
		|m|^{D-2} - \frac{D}{2} |m_0|^{D-2}
		\right)
		+ \frac{4}{(2\pi)^\frac{D}{2}}W_{\frac{D}{2}} \left[m^2;\beta,\mu,1;L,\theta\right].
	\end{multline}
	
	This result is valid for any dimensionality but in principle only applicable for $D\le3$. In 1+1 dimensions the theory is renormalizable and in 1+2 dimensions, although not perturbatively renormalizable, it was shown that the theory can be defined through the methods of constructive quantum field theory~\cite{deCalan:1991km}.
	
	\subsection{$GN_{1+1}$, Fermion on a Ring}
	For  $D=1+1$ the effective potential is 
	\begin{equation}
		V_{\text{eff}}^{D=2}=
		\frac{m^2}{4\pi} 
		\left(
		\ln \frac{m^2}{m_0^2}-1
		\right)
		+ \frac{2}{\pi}W_{1} \left[m^2;\beta,\mu,1;L,\theta\right].
	\end{equation}
	
	Its first derivative with respect to $m$ gives all extrema. Discarding the known $m=0$ result of the symmetric phase, we have the mass gap equation
	\begin{equation}
		\ln \frac{m}{m_0}
		= 2 W_{0} \left[m^2;\beta,\mu,1;L,\theta\right].
		\label{Eq:GN2_m}
	\end{equation}
	To investigate the existence of a minimal length we go directly to the critical condition $m=0$ and take a zero critical temperature $T_c=0$. We then find, after some manipulations [see Eq.~\eqref{Eq:Wpolilog0}], that
	\begin{equation}
		\ln \frac{2}{m_0 L_{\text{min}}} = \gamma + \text{Li}'_0(e^{-i \pi \theta}) + \text{Li}'_0(e^{i \pi \theta})
	\end{equation}
	
\begin{figure}[h]
		\centering
		\includegraphics[width=0.5\linewidth]{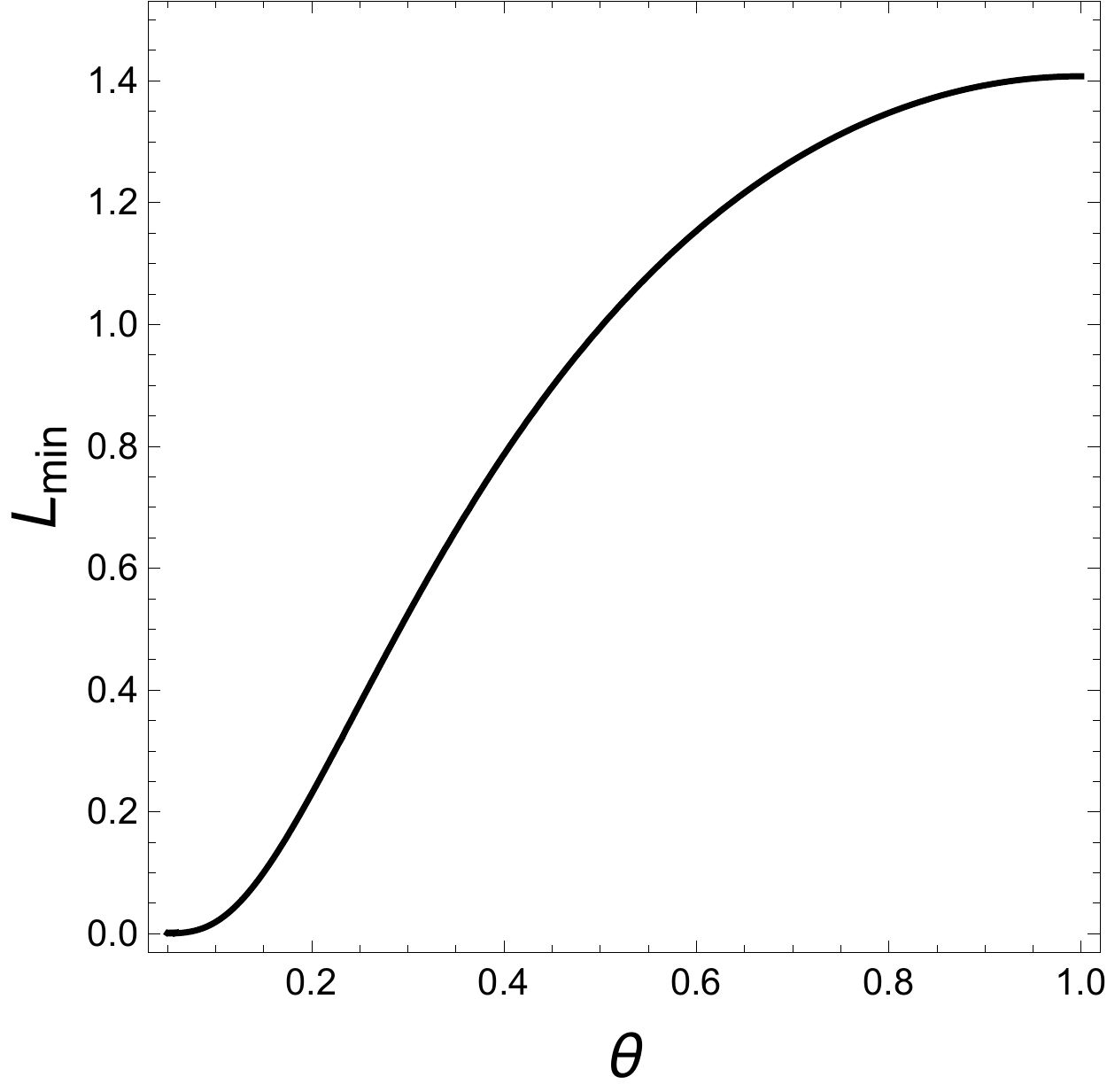}
		\caption{Minimal length as a function of the contour parameter in the Gross-Neveu model for $D=1+1$. Only for $\theta=0$ the minimal length turns out to be zero. Under the curve the phase is \textit{always} symmetric and there is no thermally induced transition}
		\label{fig:LminA}
	\end{figure}

	The minimal length is controlled by the contour parameter, and as we take lower values of $\theta$ the value of $L$ diminishes, see Fig.~\ref{fig:LminA}. It becomes zero only for $\theta^\star=0$, the fully periodic case. We must remark that this is the same result we obtained for the bosonic case: for $D=2$ ($\rho=0$) we obtain the value $\theta^\star=0$. This seems to point out a property of the formalism, independently of whether we use bosonic or fermionic models.
		
	\subsection{$GN_{1+2}$, Fermion on a Tube}
	
	As already stated, for $D=3$ the Gross-Neveu model was shown to exist and was constructed~\cite{deCalan:1991km} although it is not perturbatively renormalizable. Then, as a last example, we employ our mean-field non-perturbative approach to consider a fermion model on a tube ($D=1+2$). The effective potential is
	\begin{equation}
		V_{\text{eff}}^{D=3}=
		\frac{m^2}{12\pi}\left(2|m|-3|m_0|\right)
		+ \sqrt{\frac{2}{\pi^3}}W_{\frac{3}{2}} \left[m^2;\beta,\mu,1;L,\theta\right].
	\end{equation}
	\noindent The first derivative with respect to $m$ exhibits two solutions: a symmetric solution corresponding to $\widetilde{m}=0$ and a broken one with $\widetilde{m}\neq0$ given by
	\begin{equation}
		|\widetilde{m}| = |m_0| + \sqrt{\frac{8}{\pi}} W_{\frac{1}{2}} \left[\widetilde{m}^2;\beta,\mu,1;L,\theta\right].
	\end{equation}
	\noindent In the neighbourhood of the symmetric case, $\widetilde{m}=0$ defines a critical temperature. As the critical temperature goes to zero, $T_c=0$, a minimal length is defined,
	\begin{equation}
		|m_0| L_{\text{min}}  = \ln\left(1 - e^{i\pi \theta}\right)
		+  \ln\left(1 - e^{-i\pi \theta}\right).
	\end{equation}
	
\begin{figure}[h]
		\centering
		\includegraphics[width=0.5\linewidth]{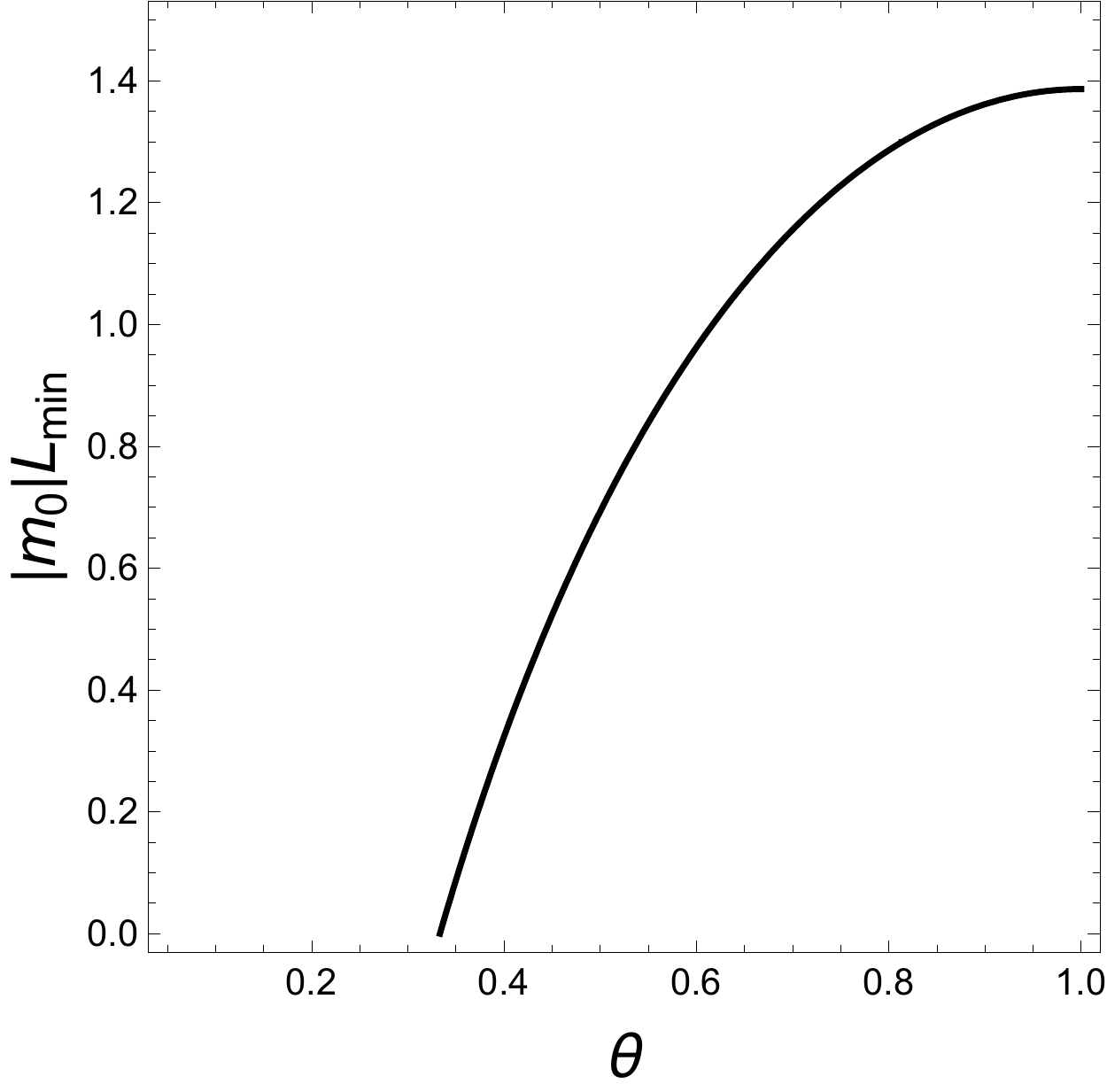}
		\caption{Minimal length the Gross-Neveu model for $D=1+2$. Under the curve the system is always in the symmetric phase.}
		\label{fig:LminD3}
	\end{figure}

	For the antiperiodic boundary condition $\theta=1$ we have the minimal length given by $|m_0| L_{\text{min}}\Big|_{a=1} = \ln 4$. Decreasing the value of the parameter $\theta$ which describes the quasiperiodic boundary condition, we find a critical contour parameter $\theta^\star=1/3$ at which the minimal length turns out to be zero, see Fig.~\ref{fig:LminD3}. This, again, is the same result we have obtained for the bosonic case when $D=3$, indicating that the critical contour is only dimensional dependent.
	
	\FloatBarrier
	\section{Discussion\label{Sec:Discussion}}

	Along this article we have assumed a contour parameter $\theta$ defining quasiperiodic boundary conditions and studied its consequences using some bosonic and fermionic models. However, we did not take into account how these different boundary conditions arise. In fact, we have emphasized that we are dealing with a mathematical aspect of the formalism and are not directly concerned with experimental comparison.
	
	We propose that the countour parameter may arise as related to a constant gauge field component along the compactified dimension~\cite{Borges1999,CcapaTtira:2010ez}. The action for a complex bosonic field minimally coupled to an external abelian gauge field is
	\begin{equation}
		S = \int d^D x \left\{ (\partial_\mu\Phi + i e A_\mu\Phi)^\star(\partial^\mu \Phi+ i e A^\mu \Phi) + m^2 \Phi^\star\Phi  \right\}.
	\end{equation}
	\noindent If we consider a constant gauge field only along the compactified dimension $x_1$ such that $A_\mu=(0,A_1,0,\ldots,0)$ where $A_1=\text{const}$, we note that this contribution is given by the substitution $p_1 \rightarrow p_1 + e A_1$. Recalling the original identification that introduced the boundary parameter, see Eq.~\eqref{Eq:Identification}, we see that the relation between $A_1$ and $\theta$ is just
	\begin{equation}
		e A_1 = \frac{\theta \pi}{L}.
	\end{equation}
	Therefore, the contour parameter can be thought of as a consequence of a constant gauge field that does not have any dependence on the Euclidean space variables. The value of $A_1$ allows interpolating between the perfect periodic and perfect antiperiodic conditions. Perhaps, this may be related to the well-known result that a constant gauge field generates an Aharonov-Bohm phase~\cite{Aharonov:1959fk}, which induces a transmutation between fermions and bosons~\cite{Polyakov:1988md}. In another context,  for which interpolation between bosons and fermions occurs in the imaginary-time variable, studies were made in which the Aharonov-Bohm phase is induced by a Chern-Simons term~\cite{Fradkin:1994tt,Dunne:1998qy}. 
	Furthermore, this topic has been the subject of a detailed study on how the relationship between the gauge field and the statistical phase emerges \cite{Forte:1990hd}. All these works~\cite{Borges1999,Aharonov:1959fk,Polyakov:1988md,Fradkin:1994tt,Dunne:1998qy,Forte:1990hd} justify the introduction of the \textit{contour parameter} $\theta$ whose consequences were studied along this article.

	It is not surprising that the contour condition (like a border effect) would influence the system even when its length is lowered to its minimal. We have exhibited, using some simple bosonic and fermionic models, that the boundary conditions directly influence the minimal length below which there is no thermally-induced transition. Furthermore, there is a critical contour parameter at which the minimal length is zero.
	
	We have found that the critical contour parameter depends only on the system dimensionality of the system. For a bosonic system, we employ two models, one with a second-order phase transition and the other with a first-order phase transition; both show the same value for the critical parameter if the dimensions are equal. We have also tested a fermionic model and find that the parameter $\theta$ has the same value. The only difference between a bosonic and a fermionic system turns out to be that for a bosonic system there is a minimal length for $\theta < \theta^\star$, while for a fermionic system there is a minimal length for $\theta  > \theta^\star$. The observed independence of $\theta^\star$ shows that there is a common substrate of models having quasiperiodic boundary conditions independent of its physical nature. 

	\section*{Acknowledgments}
	
	The authors thank the Brazilian agencies CAPES and CNPq for partial financial support. We also thank C. Farina for fruitful suggestions.
	
\bibliographystyle{ws-ijmpa}
\bibliography{AllRefs2}

\end{document}